\documentclass[prl,twocolumn,superscriptaddress,longbibliography,amsfonts,amssymb,amsmath,floatfix,floats,a4paper]{revtex4-1}
\usepackage{graphicx}
\usepackage{booktabs}
\usepackage{amsmath}
\usepackage{color}
\usepackage{hyperref}
\usepackage[breakable,skins]{tcolorbox}
\usepackage{bbold}

\definecolor{orange}{rgb}{1,0.5,0}
\definecolor{darkgreen}{rgb}{0.0, 0.578125, 0.25}
\definecolor{darkblue}{rgb}{0.0, 0, 0.7}
\definecolor{OrangeRed}{rgb}{0.92941,0.0745,0.35294}

\newcommand{\ket}[1]{\ensuremath{\left|#1\right\rangle}}
\newcommand{\bra}[1]{\ensuremath{\left\langle#1\right|}}
\newcommand{\ketbra}[2]{\ensuremath{\left|#1\right\rangle\left\langle#2\right|}}

\begin{document}
\title{Measurements of nonlocal variables and demonstration of the failure of the product rule for a pre- and postselected pair of photons}

\author{Xiao-Ye Xu}
\author{Wei-Wei Pan}
\author{Qin-Qin Wang}
\affiliation{CAS Key Laboratory of Quantum Information, University of Science and Technology of China, Hefei 230026, People's Republic of China}
\affiliation{Synergetic Innovation Center of Quantum Information and Quantum Physics, University of Science and Technology of China, Hefei 230026, People's Republic of China}
\author{Jan Dziewior}
\author{Lukas Knips}
\affiliation{Max-Planck-Institut f\"ur Quantenoptik, Hans-Kopfermann-Stra{\ss}e 1, 85748 Garching, Germany}
\affiliation{Department f\"{u}r Physik, Ludwig-Maximilians-Universit\"{a}t, 80797 M\"{u}nchen, Germany}
\author{Yaron Kedem}
\affiliation{Department of Physics, Stockholm University, AlbaNova University Center, 106 91 Stockholm, Sweden}
\author{Kai Sun}
\author{Jin-Shi Xu}
\author{Yong-Jian Han}
\author{Chuan-Feng Li}
\email{cfli@ustc.edu.cn}
\author{Guang-Can Guo}
\affiliation{CAS Key Laboratory of Quantum Information, University of Science and Technology of China, Hefei 230026, People's Republic of China}
\affiliation{Synergetic Innovation Center of Quantum Information and Quantum Physics, University of Science and Technology of China, Hefei 230026, People's Republic of China}

\author{Lev Vaidman}
\email{vaidman@post.tau.ac.il}
\affiliation{Raymond and Beverly Sackler School of Physics and Astronomy, Tel-Aviv University, Tel-Aviv 69978, Israel}

\date{\today}

\begin{abstract}
We report the first implementation of the von Neumann instantaneous measurements of nonlocal variables which becomes possible due to technological achievements in creating hyperentangled photons.
Tests of reliability and of the nondemolition property of the measurements have been performed with high precision showing the suitability of the scheme as a basic ingredient of numerous quantum information protocols.
The method allows to demonstrate for the first time with strong measurements a special feature of pre- and postselected quantum systems: the failure of the product rule.
It has been verified experimentally that for a particular pre- and postselected pair of particles a single measurement on particle $A$ yields with certainty $\sigma_x^A=-1$, a single measurement on particle $B$ yields with certainty $\sigma_y^B=-1$, and a single nonlocal measurement on particles $A$ and $B$ yields with certainty $\sigma_x^A \sigma_y^B=-1$.
\end{abstract}

\maketitle

\label{intro}

All known interactions in nature are local. 
It was thus believed (e.g.\,\cite{Landau1931}) that measurements of nonlocal variables (variables which are related to more than one region of space) are impossible. However, Aharonov and his co-authors \cite{Aharonov1980,Aharonov1986} showed theoretically that some nonlocal variables can be measured. For two separate locations  the sum of local variables, $A+B$, and the modular sum, $(A+B) \!\! \mod c$ are always measurable. 
On the other hand, they also showed that some other nonlocal variables cannot be measured as this would lead to superluminal signalling. Note that if we do not require the measurement to be nondemolition, then theoretically all nonlocal variables are measurable \cite{V2003}, but the procedure has high demands on entanglement resources \cite{entan1,entan2,entan3}.

Aharonov's main motivation was to shed light on relativistic quantum field theory \cite{Aharonov1980,AA84I,AA84II,Beck1,Ghir}, but the main impact of the analysis of measurements of nonlocal variables was in the field of quantum information \cite{VaPope94,Beck2,VGW,BennetNonL,modular2010,non1,non2,non4,non5}. 
In particular, it allowed an efficient method for teleportation \cite{V94} and was the basis for cryptographic protocols \cite{Crypto0,Crypto1,Crypto2,Crypto3}. 

In this work we demonstrate the measurement of nonlocal variables in its original sense, the one which is closest to the standard von Neumann definition of measurement in quantum mechanics \cite{Neumann}.
Note, that there exists an alternative scheme \cite{BrodCo} alongside a particular proposal for its implementation \cite{QOpt,Edam}, which, however, has the drawback of being a probabilistic measurement, i.e., even with ideal devices it might not provide an outcome.

After performing and testing our  measurement procedure we apply it  to show the peculiar phenomenon of the failure of the product rule for two separate (and thus commuting) local variables which can take place only for pre- and postselected quantum systems \cite{Hardy,MyHardyReply,HardyAharonov}.
There have been several demonstrations of the failure of the product rule for weak values, the outcomes of weak measurements \cite{HardyBowm,HardySteinberg,HardyImoto} in the context of the Hardy paradox \cite{Hardy}.
These, however, are very different results, obtained from many measurements on an ensemble of particles.
In our scenario, we are able to violate the product rule using strong nondemolition measurements, providing direct information about single pairs of particles.
We will show below that the same cannot be done in the setting of the Hardy paradox.

Let us start by spelling out the properties of a von Neumann measurement.
It has to be reliable, nondemolition and instantaneous.
  
Reliability: if the initial state is an eigenstate of the measured nonlocal variable, the corresponding eigenvalue is found with certainty.
 
Nondemolition  property: if the initial state is an eigenstate of the measured nonlocal variable, the state is not changed. 
This ensures repeatability, namely that identical consecutive measurements all agree with the reading of the first measurement.
 
Instantaneousness: The standard definition in non-relativistic quantum mechanics is that the measurement process has a negligible duration.
According to the principles of relativistic quantum mechanics there is no physical mechanism that allows a local observer simultaneous coupling to spatially separated parts of a physical system.
Thus, a measurement of a nonlocal observable must consist of two separate local couplings at distinct spatial locations followed by a creation of macroscopic records.
In addition to the requirement of negligible duration of these local couplings together with the respective creation of local records we require that the processes at the two locations are spacelike separated.
The measurement coupling and creations of macroscopic records which specify the result of the measurement happen instantaneously and simultaneously in some Lorentz frame, e.g., that of the observer.

\begin{figure}
\centering
\includegraphics[width=0.45\textwidth]{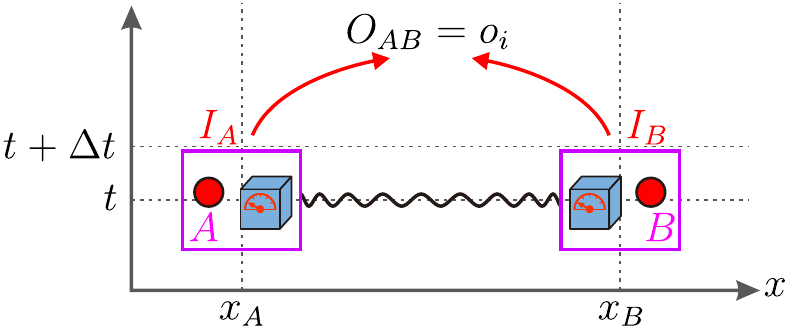}
\caption{{\bf Measurement scheme of a nonlocal variable.} Entangled parts of the measuring device in locations $A$ and $B$ couple simultaneously to the respective parts of the system for a short time. 
The locally obtained records $I_A$ and $I_B$ provide, when brought together, the eigenvalue $o_i$ of the nonlocal variable $O_{AB}$ of the system.}
\label{fig:Demon}
\end{figure}

To satisfy the above requirements our scheme for the measurement of nonlocal variables has two stages, see Fig.\,\ref{fig:Demon}. 
In the first stage, spatially separated measuring devices (``pointers'') interact locally with the respective components of the system. 
In the second stage, immediately after, macroscopic records are created via local measurements of the measuring device.  
To gain nonlocal information from these local measurements the crucial element in our scheme is the preparation of the measuring device in a nonlocal entangled state.

\begin{figure*}
\centering
\includegraphics[width=0.95\textwidth]{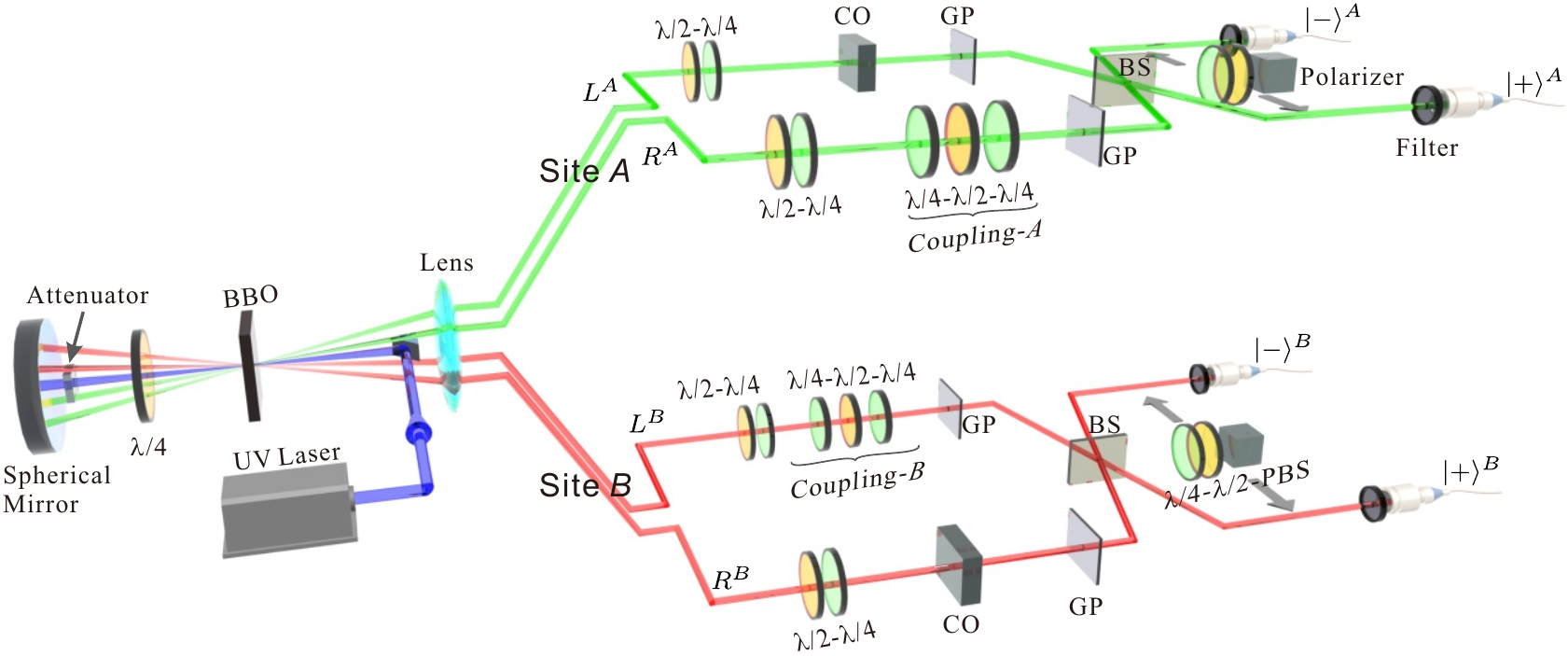}
\caption{{\bf Experimental Setup}. The two-photon path-polarization hyperentangled state is generated from the SPDC using $\beta$-BaB$_2$O$_4$ (BBO) crystal in a round-tripping confocal structure~\cite{Kwiat1997,Ciampini2016}. 
The photon $A$ will be in the upper layer (green lines), while the lower layer is site $B$ (red lines). 
The attenuator and sets of $\lambda/2$-$\lambda/4$ wave plates prepare the polarization state. 
The local coupling between the system (polarization degree of freedom) and the measuring device (path degree of freedom $L$ and $R$) is realized by sets of wave plates ($\lambda/4$-$\lambda/2$-$\lambda/4$) in a particular arm of the interferometer. For the nonlocal measurement two such local couplings are applied. In the other arms, compensation crystals (CO) ensure identical optical path length. The relative phases are tuned using thin glass plates (GP) before bringing the arms to interference using a polarization insensitive beam splitter (BS). For testing the nondemolition property polarization filters composed of set of $\lambda/4$-$\lambda/2$ plates and polarization beam splitter (PBS) are placed in front of the detectors.}
\label{fig:Setup}
\end{figure*}

We perform a measurement of the product of polarization operators of two photons separated in space.
The measuring device is given by the path degree of freedom of the photons themselves, which is measured interferometrically, see Fig.\,\ref{fig:Setup}.
In this setting, the local macroscopic records correspond to ports of the interferometer in which the detectors eventually click. 
The meaning of ``nondemolition'' is that the polarization state of the pair of photons in the corresponding pair of ports is still the initial (possibly entangled) polarization state as long as it started in an eigenstate of the nonlocal variable. 
While in our experiment there is no spacelike separation of the local measurement processes each one consisting of a coupling in one arm of the interferometer and the detector click which provides the macroscopic record, our scheme in no way relies on the proximity of the two parties and thus allows for a modification with a spacelike separation of the local measurements.

The Hilbert spaces of photon polarization states and of the states of a spin$-\frac{1}{2}$ particle are isomorphic which allows us to use a more familiar language of spin operators.
Expressed with spin operators our measurement corresponds to the measurement of $\sigma^A_z \sigma^B_z$ on two remote particles.
There is no general method for the measurement of the product of variables belonging to remote locations, but $\sigma^A_z \sigma^B_z=(\sigma^A_z + \sigma^B_z) \!\! \mod 4 -1$, so the measurement of this product is equivalent to the measurement of a modular sum which is proven to be possible \cite{Aharonov1986}.
For the polarization degree of freedom we use the correspondence
\begin{equation}\label{spin-poalrisation}
\sigma_z |{\uparrow}_z\rangle = |{\uparrow}_z\rangle \leftrightarrow |H\rangle, ~~~~  |{\downarrow}_z\rangle \leftrightarrow |V\rangle,
\end{equation}
with $\ket{H}$ and $\ket{V}$ identifying horizontal and vertical polarization, respectively. 
Similarly, for the measuring device, which we will mark with ``$\sim$'' in the spin notation,
\begin{equation}\label{spin-path}
\tilde{\sigma}_z |\tilde{\uparrow}_z\rangle  = |{\tilde{\uparrow}}_z\rangle \leftrightarrow |L\rangle, ~~~~  |{\tilde{\downarrow}}_z\rangle \leftrightarrow |R\rangle,
\end{equation}
where $\ket{L}$ and $\ket{R}$ denote the two arms the photons can take at both locations.

The measurement scheme requires to correctly match the preparation of the initial pointer state, the interaction Hamiltonian, and the pointer measurement.
We prepare the measuring device in the entangled state
\begin{equation}\label{MDstate}
|\Psi^+\rangle^{AB}\equiv\frac{1}{\sqrt 2} (|\tilde{{\uparrow}}_z\rangle^A|\tilde{{\downarrow}}_z\rangle^B + |\tilde{{\downarrow}}_z\rangle^A|\tilde{{\uparrow}}_z\rangle^B),
\end{equation}
and perform local pointer measurements in the plane perpendicular to $\tilde{\sigma}_z$. 
The interaction Hamiltonian is
\begin{equation}\label{Eq-H2}
H = g(t) [(\mathbb{1}^A - {\sigma}_z^A) (\mathbb{1}^A - \tilde{\sigma}_z^A) + (\mathbb{1}^B - {\sigma}_z^B) (\mathbb{1}^B + \tilde{\sigma}_z^B)],
\end{equation}
where $\int g(t)d t = \pi/4$. 

Given a polarization state which is an eigenstate of $\sigma_z^A \sigma_z^B$ with eigenvalue $+1$, the interaction Hamiltonian (\ref{Eq-H2}) does not change the state $|\Psi^+\rangle^{AB}$ of the measuring device.
On the other hand, for an eigenstate with eigenvalue $-1$, the state of the measuring device is changed to
\begin{align}
|\Psi^-\rangle^{AB}\equiv\frac{1}{\sqrt 2} (|\tilde{{\uparrow}}_z\rangle^A|\tilde{{\downarrow}}_z\rangle^B - |\tilde{{\downarrow}}_z\rangle^A|\tilde{{\uparrow}}_z\rangle^B).
\end{align}
The states $|\Psi^+\rangle^{AB}$ and $|\Psi^-\rangle^{AB}$ can be distinguished reliably using our simultaneous local measurements in $A$ and in $B$, after the records have been combined.
In both cases, after the interaction, the system and the pointer are not entangled with each other and the system state is not changed.
This makes the measurement nondemolition.
Note that also the product of spin operators in other directions can be measured in this way, when the Hamiltonian is adapted accordingly.
Also a strictly local spin measurement can be achieved using this scheme when only one of the two local coupling terms is present in the interaction Hamiltonian.

In our experiment (as shown in Fig.\,\ref{fig:Setup}), a vertically polarized ultraviolet ($\lambda=406.7\,\mathrm{nm}$) laser beam is focused and reflected to pump a $0.5\,\mathrm{mm}$ thick BBO crystal (type-\uppercase\expandafter{\romannumeral1} cut at $29.11^\circ$). 
Due to the degenerate spontaneous parametric down conversion (SPDC), horizontally polarized photon pairs are emitted in a cone with apex angle $3^\circ$ (postselected by spectral filters centered at $\lambda=813.4\,\mathrm{nm}$ with bandwidth $\Delta\lambda=3\,\mathrm{nm}$). 
After passing a wave plate, acting as $\lambda/4$ for $813.4\,\mathrm{nm}$ light, with optical axis oriented at $45^\circ$, both the pump laser and the emitted photons are reflected by a spherical mirror with $150\,\mathrm{mm}$ radius. 
While the polarization of previously produced photon pairs is converted to vertical, a second pass through the BBO of the pump can produce another pair of horizontally polarized photons.

The two processes overlap both spatially and temporally due to the confocal structure and the long coherence time. 
A positive lens ($f=150\,\mathrm{mm}$) transforms the conical parametric emission into a cylindrical one, preparing a so-called entanglement-ring~\cite{Ciampini2016}. 
By selecting four points in the ring, we obtain the two-photon four-qubit hyperentangled state, maximally entangled both in the polarization and spatial degrees of freedom, while separable between the two. 
The spatial state reads $|\Psi^+\rangle^{AB} = \frac{1}{\sqrt 2} (|L\rangle^A |R\rangle^B + |R\rangle^A |L\rangle^B)$.
An attenuator near the spherical mirror together with wave plate sets in the four arms of the interferometer allow to prepare arbitrary polarization states.

The two parts of the interaction Hamiltonian of Eq.~(\ref{Eq-H2}) at the two sites $A$ and $B$ are each implemented by sets of waveplates, introducing a conditional phase shift of $\pi$ for horizontally polarized photons in the right arm at site $A$ and in the left arm at site $B$.
Using glass plates we tune the interferometer such that without interaction we get complete correlation between clicks of the detectors in sites $A$ and $B$.
Then, observing the clicks corresponds to a ``spin'' measurement in some unknown direction in the $x-y$ plane, which, however, is the same for the two particles.
In the notation of the photon states, these are the measurements in the basis
 $|\pm\rangle\equiv \frac{1}{\sqrt 2} (|L\rangle \pm e^{i\varphi}|R\rangle)$ with unknown phase $\varphi$, which is equal at both sites.
In this notation, the states of the measuring device (3) and (5) are
\begin{subequations}
\begin{align}
 \ket{\Psi^+}^{AB} &= \frac{e^{-2i\varphi}}{\sqrt{2}} \big( \ket{++}^{AB} - \ket{--}^{AB} \big), \\
 \ket{\Psi^-}^{AB} &= \frac{e^{-2i\varphi}}{\sqrt{2}} \big( \ket{+-}^{AB} - \ket{-+}^{AB} \big),
\end{align}
\end{subequations}
where we used the shortcut notation ${\ket{++}^{AB}\equiv \ket{+}^{A}\ket{+}^{B}}$, etc. Thus, the  correlations and the anticorrelations of the outcomes for the measurements of $\ket{+}$ and $\ket{-}$ in the path degree of freedom allow to distinguish between the eigenstates $+1$ and $-1$ of the product operator of the system.

\begin{figure}[ht]
\centering
\includegraphics[width=0.48\textwidth]{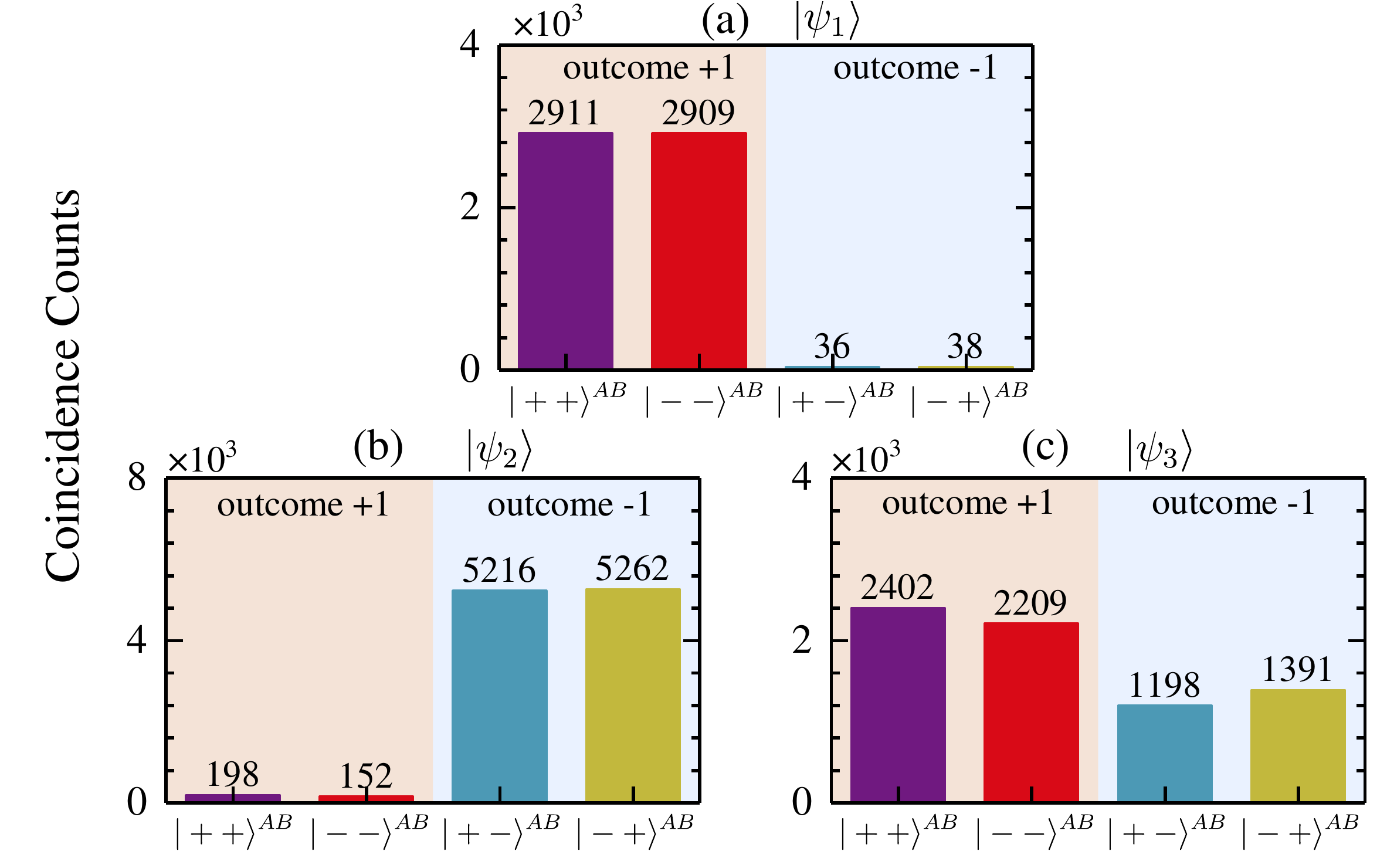}
\caption{{\bf Reliability test.} Coincidence counts on various pairs of detectors for the three polarization states of Eq.~(\ref{eq:polarizationStates}). 
The states $\ket{\psi_1}$ and $\ket{\psi_2}$ are eigenstates of the nonlocal variable $\sigma_z^A\sigma_z^B$.
The percentage of outcomes for state $\ket{\psi_3}$ corresponds well to the initial superposition.}\label{tab:reliability}
\end{figure}

We test the reliability and the nondemolition property of our measurement procedure for three different states,
\begin{subequations}\label{eq:polarizationStates}
\begin{align}
    \ket{\psi_1}&=\ket{\uparrow_z}^A\ket{\uparrow_z}^B,\\
    \ket{\psi_2}&=\frac{1}{\sqrt 2}\left(\ket{\uparrow_z}^A\ket{\downarrow_z}^B+ i\ket{\downarrow_z}^A\ket{\uparrow_z}^B\right),\\ 
    \ket{\psi_3}&=\sqrt{0.5}\ket{\uparrow_z}^A\ket{\uparrow_z}^B+\sqrt{0.1}\ket{\uparrow_z}^A\ket{\downarrow_z}^B \nonumber\\
    &-i\sqrt{0.2}\ket{\downarrow_z}^A\ket{\uparrow_z}^B+\sqrt{0.2}\ket{\downarrow_z}^A\ket{\downarrow_z}^B.
\end{align}
\end{subequations}
The results of the reliability test are presented in Fig.~\ref{tab:reliability}.
For $\ket{\psi_1}$ and $\ket{\psi_2}$ only $1.3\,\%$ and $3.2\,\%$ of the events are erroneous, respectively. State
$\ket{\psi_3}$, which is not an eigenstate of the nonlocal operator, leads to $64\,\%$ events corresponding to outcome $+1$ and $36\,\%$ to outcome $-1$ roughly agreeing with the theoretical predictions of $70\,\%$ and $30\,\%$, respectively.
 
\begin{figure}
\centering
\includegraphics[width=0.48\textwidth]{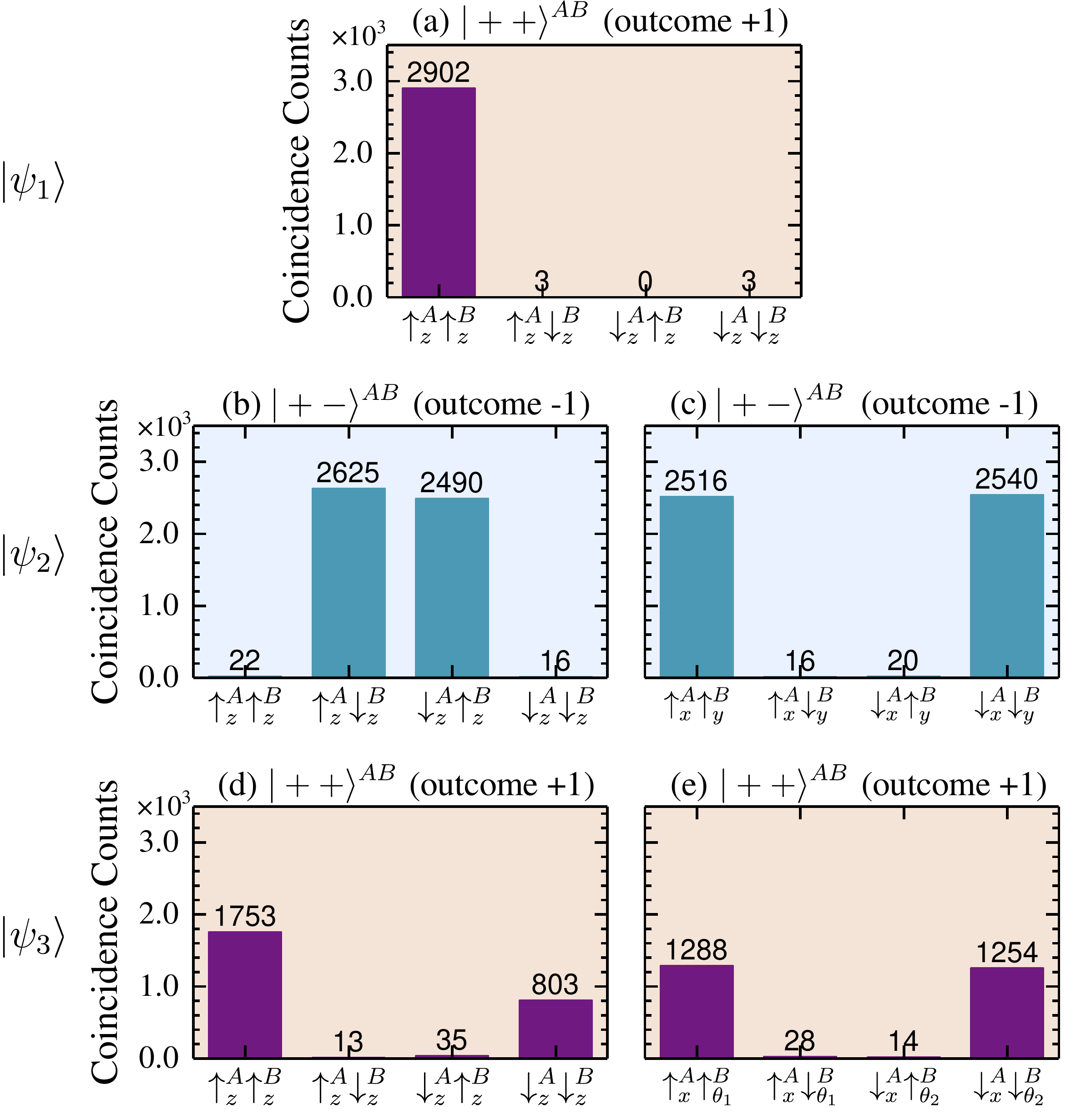}
\caption{{\bf Test of nondemolition property}. 
(a) $\ket{\psi_1}$ is measured on the detectors $\ket{++}^{AB}$ in the $\sigma^A_z,\sigma^B_z$ basis using polarization filters to verify that the state is unaltered during the coupling.
 (b,c) For  $\ket{\psi_2}$, the measurements were done on detectors $\ket{+-}^{AB}$ in bases  $\sigma^A_z,\sigma^B_z$ and  $\sigma^A_x,\sigma^B_y$.
(d) State $\ket{\psi_3}$ was measured first in  $\sigma^A_z,\sigma^B_z$ basis.
To fix the phase relation between $\ket{\uparrow_z^A}\ket{\uparrow_z^B}$ and $\ket{\downarrow_z^A}\ket{\downarrow_z^B}$, a second measurement (e) was conducted with projections containing also rotated states with $\ket{\uparrow_{\theta_{1}}}=(\sqrt{0.5}\ket{\uparrow_z} + \sqrt{0.2}\ket{\downarrow_z})/\sqrt{0.7}$ and $\ket{\uparrow_{\theta_{2}}}=(\sqrt{0.5}\ket{\uparrow_z} - \sqrt{0.2}\ket{\downarrow_z})/\sqrt{0.7}$. 
The results prove that we performed nondemolition measurements of nonlocal variables.
}
\label{fig:nondemolition}
\end{figure}

\begin{figure}
\centering
\includegraphics[width=0.48\textwidth]{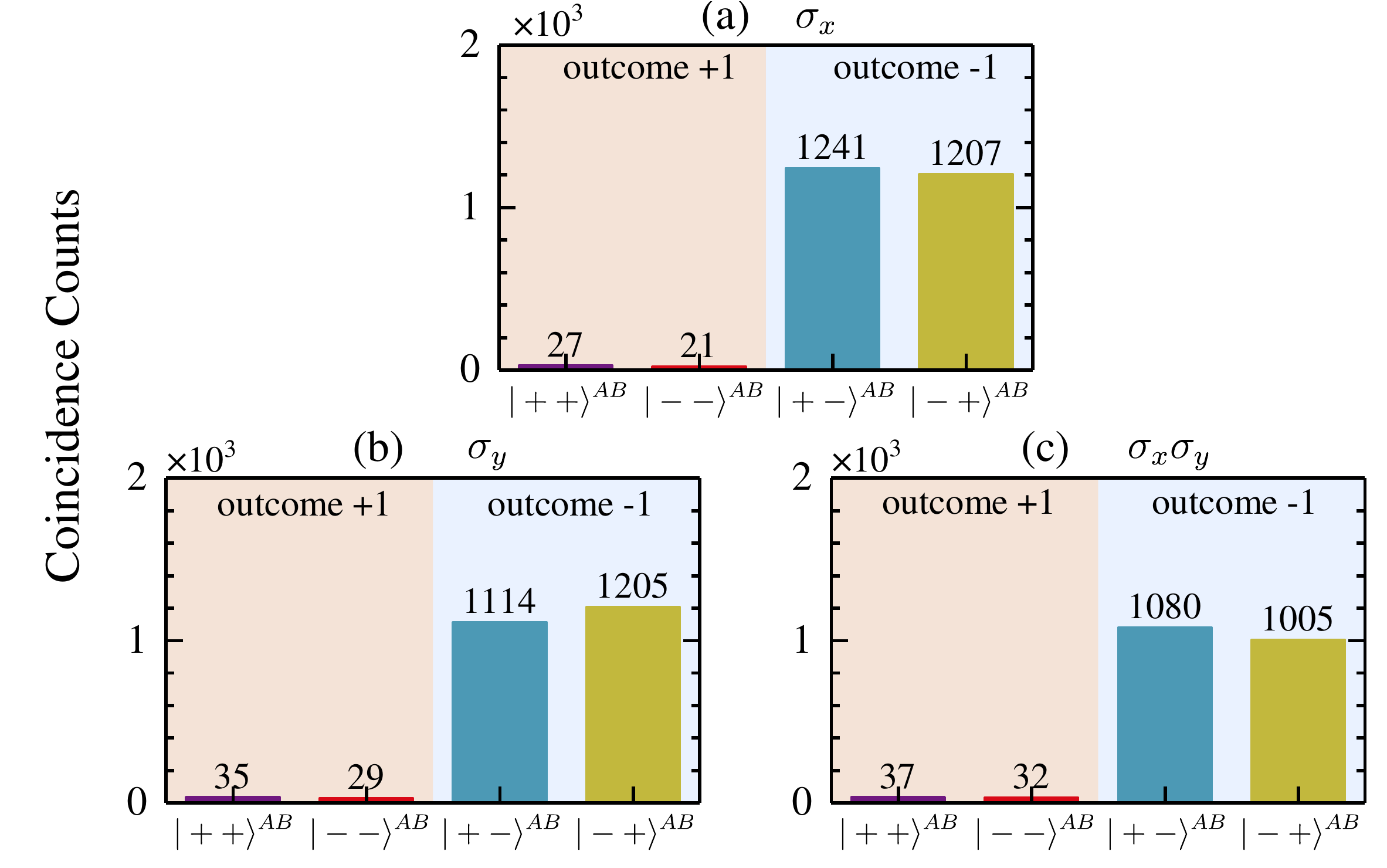}
\caption{{\bf Experimental results for demonstrating the failure of the product rule}. Coincidence counts in pairs of detectors for measurement of local and nonlocal variables: (a) $\sigma_x^A$,  (b) $\sigma_y^B$,  (c) $\sigma_x^A\sigma_y^B$. In all three cases, the system is preselected in a singlet state and postselected in the product state $\ket{\uparrow_y^A\downarrow_x^B}$.} 
\label{fig:ProductRule}
\end{figure}

To confirm that the measurement is nondemolition, we perform state verification measurements by adding polarization analyzers directly in front of the detectors. 
We test the polarization in one of the pairs of output ports, between which coincidences are expected, and repeat the test for the same measurement time with differently set filters. For the state $\ket{\psi_1}$ the results are very robust, as shown in Fig.\,\ref{fig:nondemolition}(a), as only $0.2\,\%$ changed their polarization. This measurement verifies that the output state is indeed $\ket{\uparrow_z^A}\ket{\uparrow_z^B}$.
The entangled state $\ket{\psi_2}$ requires two tests.
The first measurement in the basis $\sigma_z^A,\sigma_z^B$ shows that the state corresponding to the $-1$ outcome ($\ket{+-}^ {AB}$) of $\ket{\psi_2}$ is in a superposition (or mixture) of $\ket{\uparrow_z^A}\ket{\downarrow_z^B}$ and $\ket{\downarrow_z^A}\ket{\uparrow_z^B}$ according to Fig.\,\ref{fig:nondemolition}(b).
The second measurement in basis $\sigma_x^A$, $\sigma_y^B$ (with $\sigma_x\equiv\ket{\downarrow_z}\bra{\uparrow_z}+\ket{\uparrow_z}\bra{\downarrow_z}$ and $\sigma_y\equiv i\ket{\downarrow_z}\bra{\uparrow_z}-i\ket{\uparrow_z}\bra{\downarrow_z}$), see Fig.\,\ref{fig:nondemolition}(c), fixes the phase relation between those contributions.
In both bases, we get only $0.7\,\%$ errors. 
For state $\ket{\psi_3}$, we test the projection onto the subspace corresponding to outcome $+1$ in the ports $\ket{++}^{AB}$.
Again, we observe a good correspondence with theoretical predictions, see Fig.\,\ref{fig:nondemolition}(d) and (e).

In the second part of the experiment we use our technique of measurement of nonlocal variables to demonstrate the failure of a product rule for pre- and postselected quantum systems.

Before presenting our measurement we show that the failure of the product rule appearing in the Hardy paradox \cite{Hardy} (which was demonstarted using weak measurements \cite{HardyBowm,HardySteinberg,HardyImoto}) cannot be observed using von Neumann measurement.
Such a measurement would allow superluminal communication between Alice at site $A$ and Bob at remote site $B$.
In this case Alice would be able to change the probability distribution at Bob's site instantaneously.

Assume that a nondemolition measurement of ${\rm \bf P}^A_1 {\rm \bf P}^B_1$ is performed at time $t$ with ${\rm \bf P}^A_1 \equiv \ketbra{1}{1}^A$ and ${\rm \bf P}^B_1 \equiv \ketbra{1}{1}^B$.
Just before $t$, Bob prepares the state $\frac{1}{\sqrt 2}\left(\ket{0}^B+\ket{1}^B\right)$ and performs a projection measurement of this state immediately after time $t$.
If, also just before $t$, Alice prepares state $\ket{0}^A$, then at time $t$ the state in the two sites is an eigenstate of ${\rm \bf P}^A_1 {\rm \bf P}^B_1$ with eigenvalue 0, so the nonlocal measurement will not change the state and  Bob will find  $\frac{1}{\sqrt 2}\left(\ket{0}^B+\ket{1}^B\right)$ with probability 1.
If, instead,  Alice, just before $t$, prepares state $\ket{1}^A$, the nonlocal measurement will either lead to state $\ket{1}^A\ket{0}^B $ for the outcome ${\rm \bf P}^A_1 {\rm \bf P}^B_1=0$ or to state $\ket{1}^A \ket{1}^B $ for the outcome ${\rm \bf P}^A_1 {\rm \bf P}^B_1=1$.
In both cases the probability for Bob to find the state $\frac{1}{\sqrt 2}\left(\ket{0}^B+\ket{1}^B\right)$ is $\frac{1}{2}$ and Alice would have changed the probability distribution of Bob's outcomes instantaneously by her state preparation.

For our demonstration of the failure of the product rule we measure the product $\sigma_x^A \sigma_y^B $, instead of measuring a product of projection operators,  which {\it is} measurable in von Neumann sense.
In our example we have a pair of particles for which we know with certainty that a single local measurement of $\sigma_x^A$ or of $\sigma^B_y$ results in the outcome $-1$.
If, however, we measure the product $\sigma_x^A \sigma_y^B$ it will also provide the outcome $-1$ with certainty, instead of the naively expected product outcome $+1$.

To this end we preselect the singlet state $\frac{1}{\sqrt{2}}(\ket{\uparrow_z}^A\ket{\downarrow_z}^B- \ket{\downarrow_z}^A\ket{\uparrow_z}^B)$ and postselect  the product state $\ket{\uparrow_y}^A\ket{\uparrow_x}^B$ \cite{MyHardyReply}.
Indeed, the initial state is anticorrelated for all directions, so $\sigma_x^A=-1$ follows immediately from postselection of $\ket{\uparrow_x}^B$ and  $\sigma_y^B=-1$ follows immediately from postselection of $\ket{\uparrow_y}^A$.
We can see that $\sigma^A_x \sigma^B_y = -1$ with certainty by noticing that the outcome $+1$ can never occur.
If prior to the postselection the nonlocal measurement yields $\sigma^A_x \sigma^B_y = +1$, the initial state is projected on the space spanned by states $\ket{\uparrow_x}^A\ket{\uparrow_y}^B$ and $\ket{\downarrow_x}^A\ket{\downarrow_y}^B$.
Then, in the basis of $\sigma^A_y$ and $\sigma^B_x$ the projected state is $-\frac{1+i}{2}\left(\ket{\uparrow_y}^A\ket{\downarrow_x}^B + \ket{\downarrow_y}^A\ket{\uparrow_x}^B\right)$.
Since it is orthogonal to the postselected state $\ket{\uparrow_y}^A\ket{\uparrow_x}^B$, the postselection will never be successful and thus the outcome of the measurement will always be $\sigma_x^A \sigma_y^B=-1$.

To implement this measurement of $\sigma_x^A \sigma_y^B $ we only need a simple modification of our measurement scheme. The interaction Hamiltonian is changed to
\begin{equation}\label{Eq-H3}
H = g(t) [(\mathbb{1}^A-{\sigma}_x^A)(\mathbb{1}^A-\tilde{\sigma}_z^A) + (\mathbb{1}^B-{\sigma}_y^B)(\mathbb{1}^B+\tilde{\sigma}_z^B)],
\end{equation}
which is achieved experimentally in a straightforward way by modifying the settings  of the wave plates  at the two sites accordingly.
The measurement procedure is then completely analogous to the measurement of $\sigma^A_z \sigma^B_z$.
For the local measurements of $\sigma_x^A$ and $\sigma_y^B$, the coupling at the respective opposite site is switched off by setting the optical axis of the wave plates at zero, such that the interaction Hamiltonians become ${H = g(t)  (\mathbb{1}^A-{\sigma}_x^A)(\mathbb{1}^A-\tilde{\sigma}_z^A)}$ and ${H = g(t) (\mathbb{1}^A-\tilde{\sigma}_z^A)(\mathbb{1}^B+\tilde{\sigma}_z^B)}$.

The results are presented in Fig.\,\ref{fig:ProductRule}.
As in the nonlocal measurement of $\sigma_z^A \sigma_z^B$ explained above, the correlations and the anticorrelations of the outcomes for the measurements of $\ket{+}$ and $\ket{-}$ in the path degree of freedom allow to distinguish between the eigenstates $+1$ and $-1$ of the product operator of the system.
Fig.\,\ref{fig:ProductRule}a shows $\sigma_x^A \sigma_y^B =-1$ with very high precision.
For the local measurement at site $A$ ($B$) we use the same procedure without the interaction at site $B$ ($A$) such that again, the observed anticorrelations, Fig.\,\ref{fig:ProductRule}b,c , clearly demonstrate $\sigma_x^A  =-1$ and $ \sigma_y^B =-1$.

We have performed the first experimental implementation of von Neumann measurements of nonlocal variables.
We also demonstrated a peculiar property of pre- and postselected composite systems: the failure of the product rule for commuting observables.
The experiment sheds light on the fundamental question of the physical meaning of nonlocal variables. 
The high fidelity of the experimental results demonstrates that the method is potentially useful for  quantum information applications.

\begin{acknowledgments}
This work was supported by National Key Research and Development Program of China (Nos.\,2017YFA0304100, 2016YFA0302700), the National Natural Science Foundation of China (Nos.\,11474267, 61327901, 11774335, 61322506, 11821404), Key Research Program of Frontier Sciences, CAS (No.\,QYZDY-SSW-SLH003), Anhui Initiative in Quantum Information Technologies (AHY020100, AHY060300), the Fundamental Research Funds for the Central Universities (No.\,WK2470000026), the National Postdoctoral Program for Innovative Talents (No.\,BX201600146), China Postdoctoral Science Foundation (No.\,2017M612073), and the German-Israeli Foundation for Scientific Research and Development Grant No. I-1275-303.14. 
J.D. and L.K. gratefully appreciate support by the international Max-Planck-Research school for Quantum Science and Technology (IMPRS-QST) and the international PhD program ExQM from the Elite Network of Bavaria, respectively.
\end{acknowledgments}


\end{document}